\documentclass[journal=ancac3,manuscript=article]{achemso}

\usepackage[utf8]{inputenc}
\usepackage[T1]{fontenc}       % Use modern font encodings
\usepackage{mathtools}
\usepackage{chemformula}

\author{Stefano~{Dal~Forno}}
\email{tenobaldi@gmail.com}
\affiliation{Department of Physics, Imperial College London, London SW7 2AZ, United Kingdom.}
\author{Johannes~Lischner}
\email{j.lischner@imperial.ac.uk}
\affiliation{Department of Physics and Department of Materials, and the Thomas Young Centre for Theory and Simulation of Materials, Imperial College London, London SW7 2AZ, United Kingdom}

\title[Hot electron thermalization in TiN]{Electron-phonon coupling and hot electron thermalization  in titanium nitride}


\begin{document}

%\begin{tocentry}
%\includegraphics{./images/toc.pdf}
%\end{tocentry}

\begin{abstract}

We have studied the thermalization of hot carriers in both pristine and defective titanium nitride (\ch{TiN}) using a two-temperature model. All parameters of this model, including the electron-phonon coupling parameter, were obtained from first-principles density-functional theory calculations. The virtual crystal approximation was used to describe defective systems. We find that thermalization of hot carriers occurs on much faster time scales than in gold as a consequence of the significantly stronger electron-phonon coupling in \ch{TiN}. Specifically, the largest thermalization times, on the order of 200 femtoseconds, are found in \ch{TiN} with nitrogen vacancies for electron temperatures around 4000 K. 
\end{abstract}

\section{Introduction}

Titanium nitride (TiN) is a metal with a high melting point and good electrical and thermal conductivities \cite{kuwahara_mechanical_2001, andrievski_physical-mechanical_1997, burnett_mechanical_1987, taylor_thermal_1964, hojo_defect_1977}. It has been used in a number of industrial applications as a coating material because of its favorable mechanical and thermal properties \cite{pierson_handbook_2013}. Recently, \ch{TiN} has attracted significant interest as a cheap material for plasmonic applications because of its attractive optical properties~\cite{patsalas_optical_2001,patsalas_optical_2015}. Specifically, \ch{TiN} can sustain very high power illumination without melting and hence it is a promising material for hot-carrier devices \cite{murai_plasmonic_2016,naik_titanium_2012,ishii_titanium_2016}.

The dynamics of photoexcited electrons and holes in \ch{TiN} has been studied using pump-probe spectroscopy~\cite{doiron_optimizing_2018,doiron_plasmon-enhanced_2019,anisimov_electron_1974,elsayed-ali_time-resolved_1987,elsayed-ali_femtosecond_1991,del_fatti_nonequilibrium_2000}. In this technique, the distribution of electrons is driven out of equilibrium via the absorption of photons (pump pulse), while the phonons initially retain their equilibrium occupancies. The resulting highly non-thermal population of electrons equilibrates within 10 -- 100 femtoseconds due to electron-electron scattering. At later times on the scale of a few picoseconds, the interaction with phonons becomes the main thermalization channel and leads to the cooling of the electrons. Eventually, the electron and phonon systems reach an equilibrium temperature and heat diffusion returns the system to its initial state.

A specific challenge in understanding hot-carrier thermalization in \ch{TiN} is the role of defects. \ch{TiN} is known to be a highly non-stoichiometric material with oxygen defects and nitrogen vacancies being the most common defects of real \ch{TiN} samples \cite{chen_oxidation_2005,toth_transition_2014,spengler_raman_1978,yuan_new_2013}. Defective \ch{TiN} can crystallize in a number of different structures depending on the defects concentration \cite{schoen_band_1969,herzig_vacancy_1987,barman_electronic_1994}.

First-principles calculations can give detailed insights into the properties of electrons, phonons and their interactions in TiN \cite{yu_phase_2015,zhu_phase_2014,yang_investigations_2016}. Ab initio density functional theory (DFT) calculations have been carried out to investigate the optical and plasmonic properties of bulk and surface \ch{TiN} \cite{mehmood_electronic_2015, marlo_density-functional_2000,wang_surface_2010,marksteiner_electronic_1986,ishii_titanium_2016}. Recently, Habib, Florio and Sundararaman studied electron-phonon interactions and hot-carrier thermalization of various transition metal nitrides, but only considered defect-free systems~\cite{habib_hot_2018}.

In this paper, we study the thermalization of photo-excited electrons due to interactions with phonons in both pristine and defective \ch{TiN} from first principles.
In particular, we calculate properties of electrons and phonons as well as electron-phonon coupling strengths and use these results to determine the material-specific parameters of a two temperature model (2TM). Defects, such as nitrogen vacancies and oxygen substitutionals, are described via the virtual crystal approximation (VCA) \cite{parmenter_energy_1955}. This allows us to make detailed predictions for the electron thermalization times in pristine and defective \ch{TiN} as function of the illumination intensity. We find that electron thermalization in \ch{TiN} occurs on time scales of less than a picosecond -- significantly faster than in gold which is often used for plasmonic applications.

\section{Methods}

In recent years, a variety of methods have been developed to describe the ultrafast dynamics of electrons and nuclei in materials from first principles \cite{campillo_hole_2000,volkov_attosecond_2019,bernardi_theory_2015}. Ab initio time-dependent density-functional theory has been employed to explain pump-probe experiments in molecules and solids \cite{faber_electronphonon_2012,neidel_probing_2013,giovannini_simulating_2013}.
This method describes the dynamics of electrons with high accuracy, but approximations, such as the Ehrenfest approach, are required to deal with the nuclear dynamics \cite{baskov_improved_2019}. Alternatively, semiclassical approaches based on the Boltzmann equation - sometimes with material-specific parameters from first-principles calculations - have been used \cite{pizzi_boltzwann:_2014,li_electrical_2015,bernardi_ab_2014,jhalani_ultrafast_2017}. Frequently, however, experimental pump-probe experiments are analyzed using the simpler two-temperature model (2TM), which can be derived from the Boltzmann equation \cite{allen_theory_1987}. Owing to its simplicity, the 2TM can be applied to complex systems, such as nanoparticles or disordered materials, which are challenging to model with ab initio techniques \cite{carpene_ultrafast_2006, chen_semiclassical_2006,jiang_improved_2005}.

Here, we employ the 2TM to describe the ultrafast thermalization of photoexcited electrons in TiN due to interactions with phonons. In this approach, the time evolution of the electron and phonon temperatures, $T_e$ and $T_p$, is determined by the two coupled differential equations
\begin{equation}
\label{eq:2TM1}
\begin{aligned}
C_{e}(T_e) \frac{dT_e}{dt} & = \nabla \cdot (\kappa_e \nabla T_e) - G(T_e,T_p) \times (T_e - T_p) + S(t), \\
C_{p}(T_p) \frac{dT_p}{dt} & = \nabla \cdot (\kappa_p \nabla T_p) + G(T_e,T_p) \times (T_e - T_p).
\end{aligned}
\end{equation}
Here, $C_e$ and $C_p$ denote the electron and phonon heat capacities, $\kappa_e$ and $\kappa_p$ are the electron and phonon thermal conductivities, $S$ is the external power source and $G$ is the electron-phonon coupling factor, which in general depends on both the electron and phonon temperatures.

Application of the 2TM to TiN nanostructures, such as thin films, allows for certain simplifications. First, we note that the terms containing the thermal conductivities can be safely neglected because the temperature of the nanostructure will quickly become spatially uniform~\cite{guler_local_2013,costescu_thermal_2003}. Also, instead of choosing an explicit form for $S(t)$, the following boundary conditions are chosen: we assume the initial electron temperature $T_e(t=0)$ to be the temperature of the hot electron gas after illumination by the pump pulse and thermalization due to electron-electron interactions.

With these assumptions, we can subtract the two equations and obtain
\begin{equation}
\label{eq:2TM2}
\frac{d}{dt}(T_e-T_p) = - G(T_e,T_p)\left( \frac{1}{C_e(T_e)}+\frac{1}{C_p(T_p)}\right) \times (T_e-T_p).
\end{equation}
In general, Eq.~\eqref{eq:2TM2} cannot be solved analytically since $G$, $C_e$ and $C_p$ depend on the electron or phonon temperatures. However, if this dependency is ignored, the electron temperature is found to be
\begin{equation}
\label{eq:2TM3}
T_e(t) = T_e^0 e^{-t/\tau_{ep}} + T_e^{\infty},
\end{equation}
where $T_e^0=T_e(t=0)$ is the initial electron temperature and $T_e^{\infty}$ is the temperature of the material when equilibrium has been reached. In the above equation, the hot electron thermalization time due to electron-phonon scattering has been defined as
\begin{equation}
\label{eq:2TM4}
\frac{1}{\tau_{ep}} = G(T_e,T_p)\left( \frac{1}{C_e(T_e)}+\frac{1}{C_p(T_p)}\right).
\end{equation}
We have verified through explicit numerical solution of Eq.~\eqref{eq:2TM2} that the analytical expression for $\tau_{ep}$ yields accurate results for TiN.

To obtain the value of $\tau_{ep}$ for TiN, we determine all material-specific parameters from first-principles calculations based on density-functional theory~ \cite{brown_ab_2016}.
Specifically, the electron and phonon heat capacities are given by
\begin{equation}
\label{eq:elheat}
C_e(T_e) = \int_{-\infty}^{+\infty} \epsilon g_{e}(\epsilon) \frac{\partial f(\epsilon, T_e)}{\partial T_e} d \epsilon,
\end{equation}
\begin{equation}
\label{eq:phheat}
C_p(T_p) = \int_{0}^{+\infty} \epsilon g_{p}(\epsilon) \frac{\partial n(\epsilon, T_p)}{\partial T_p} d \epsilon,
\end{equation}
where $g_e(\epsilon)$, $g_p(\epsilon)$, $f(\epsilon, T_e)$ and $n(\epsilon, T_p)$ are the electron and phonon density of states and the Fermi-Dirac and Bose-Einstein distributions, respectively. We do not take the temperature dependence of the density of states into account which is an excellent approximation $T < 10^4$~K \cite{zijlstra_modeling_2013,vinko_electronic_2010,recoules_effect_2006}. Also, we set $T_p=300$~K since we are interested in processes which are much faster than the heat diffusion by phonons.

The electron-phonon coupling parameter is given by
\begin{equation}
\label{eq:gparam}
G(T_e) = - \frac{\pi k_B}{\hbar g_e(E_F)} \lambda \langle \omega^2 \rangle \int_{-\infty}^{+\infty} g^2_{e}(\epsilon) \frac{\partial f(\epsilon, T_e)}{\partial T_e} d \epsilon,
\end{equation}
where $g_e(E_F)$ is the electron density of states at the Fermi level and $\lambda \langle \omega^2 \rangle$ is the second moment of the Eliashberg function (defined below). Eq.~\eqref{eq:gparam} rests on two approximations. First, the momentum-dependent electron-phonon matrix elements have been averaged over the first Brillouin zone (isotropic approximation). Second, the phonon energies are assumed to be small compared to $k_B T_e$ and hence only transitions occurring near the Fermi surface are taken into account. As a result, $G$ is independent of the phonon temperature and can be computed straightforwardly from first principles.
We further neglect non-equilibrium effects in $G$~\cite{mueller_nonequilibrium_2014}.

At low electron temperatures, the expression for $G$ can be further simplified to yield
\begin{equation}
\label{eq:gapprox}
G \propto g_e(E_F) \lambda \langle \omega^2 \rangle.
\end{equation}
Inserting this expression into 
Eq.~\ref{eq:2TM4} yields 
\begin{equation}
\label{eq:allen}
\tau_{ep} = \frac{\pi k_B T_e}{3 \hbar \lambda \langle \omega^2 \rangle}.
\end{equation}
This important result for the electron thermalization time in metals due to phonon scattering was first obtain by Allen~\cite{allen_theory_1987}.

Finally, we define the isotropic Eliashberg function, its $n$-th moment and the phonon linewidths as 
\begin{equation}
\label{eq:eliashberg}
\alpha^2 F(\omega) = \frac{1}{g_e(E_F)} \int \frac{d\mathbf{k} d\mathbf{q}}{\Omega^2_{BZ}} \sum_{nm\nu} |g_{nm\nu}(\mathbf{k},\mathbf{q})|^2
\delta(\epsilon_{n\mathbf{k}}-E_F) 
\delta(\epsilon_{m\mathbf{k}+\mathbf{q}}-E_F)
\delta(\hbar \omega_{\mathbf{q}\nu} - \hbar\omega),
\end{equation}
\begin{equation}
\label{eq:lambdaomega}
\lambda \langle \omega^n \rangle = 2 \int_0^{+\infty} \alpha^2 F(\omega) \omega^{n-1} d\omega,
\end{equation}
\begin{equation}
\label{eq:gamma}
\gamma_{\mathbf{q}\nu} = 2 \pi \omega_{\mathbf{q}\nu} \int \frac{d\mathbf{k}}{\Omega_{BZ}} \sum_{nm} |g_{nm\nu}(\mathbf{k},\mathbf{q})|^2 \delta(\epsilon_{n\mathbf{k}}-E_F) 
\delta(\epsilon_{m\mathbf{k}+\mathbf{q}}-E_F),
\end{equation}
where $|g_{nm\nu}(\mathbf{k},\mathbf{q})|^2$ are the electron-phonon matrix elements of an electronic transition from state $\{n\mathbf{k}\}$ to state $\{m\mathbf{k}+\mathbf{q}\}$ induced by a phonon of frequency $\omega_{\mathbf{q}\nu}$ and $\Omega_{BZ}$ is the volume of the first Brillouin zone. Eq.~\eqref{eq:gamma} relies on the double-delta approximation valid for $k_b T \gg \omega_{\mathbf{q}\nu}$.
Equations~\eqref{eq:eliashberg}, \eqref{eq:lambdaomega} and \eqref{eq:gamma} reveal an interesting connection between phonon linewidths (which are typically measured in vibrational spectroscopy experiments) and electron relaxation times (which are measured in pump-probe experiments). In particular, $\lambda \langle \omega^2 \rangle$ which determines the electron relaxation time can be expressed as a sum over phonon linewidths. Thus, analysis of the phonon linewidths allows us to identify the most important relaxation channels that contribute to electron relaxation. Further details on the Green's function formalism used to obtain Eqs.~\eqref{eq:eliashberg}, \eqref{eq:lambdaomega}, \eqref{eq:gamma} can be found in the review by Giustino \cite{giustino_electron-phonon_2017}.

\subsection{Computational details}

We performed density functional theory (DFT) calculations of the electronic states, phonons and electron-phonon matrix elements of TiN using the Quantum Espresso software package~\cite{giannozzi_quantum_2009}. We focus on the dominant rock salt structure and consider both pristine and defective systems. Specifically, we carry out calculations for nitrogen-deficient \ch{TiN_x} (0.84 $\le$ x < 1.00) and also for systems with oxygen substitutional defects \ch{TiN_xO_{1-x}} (0.80 $\le$ x < 1.00). The following stoichiometries were investigated: \ch{TiN}, \ch{TiN_{0.80}O_{0.20}}, \ch{TiN_{0.85}O_{0.15}}, \ch{TiN_{0.90}O_{0.10}}, \ch{TiN_{0.95}O_{0.05}}, \ch{TiN_{1.00}}, \ch{TiN_{0.97}}, \ch{TiN_{0.94}}, \ch{TiN_{0.91}}, \ch{TiN_{0.88}} and \ch{TiN_{0.84}}. To model defective \ch{TiN} systems we used the virtual crystal approximation (VCA) as explicit electron-phonon calculations on supercells including defects are numerically prohibitive. The VCA captures average changes in the atomic mass, but also electronic effects, such as changes in the Fermi level, the density of states \cite{yu_efficient_2007,iniguez_first-principles_2003,jong_influence_2016}and also electron-phonon interaction in complex materials \cite{pena-seaman_first-principles_2007,de_la_pena-seaman_effects_2009}. We note, however, that the VCA does not capture all defect-induced relaxation channels and therefore the calculated relaxation times should be interpreted as upper bounds to the experimentally measured ones.

We employed ultra-soft pseudopotentials and the BLYP exchange-correlation functional~\cite{vanderbilt_soft_1990,garrity_pseudopotentials_2014}. We use wavefunction and charge density cutoffs of 60~Ry and 600~Ry, respectively, and a $24 \times 24 \times 24$ k-point mesh. The Marzari-Vanderbilt cold smearing approach is used for the occupancies with a broadening of 0.02~Ry~\cite{marzari_thermal_1999}. For comparison, we also present results for Au where hot electron thermalization has been extensively studied and is well understood~\cite{sun_femtosecond-tunable_1994,suarez_dynamics_1995,mueller_nonequilibrium_2014,brown_ab_2016,jain_thermal_2016}.

For each system, we first determine the relaxed unit cell size. For pristine \ch{TiN} we find $a=8.11$ Bohr, in good agreement with the experimental value $a_{expt}=8.01$ Bohr~\cite{herzig_vacancy_1987}. For the relaxed systems, we obtain phonon properties using a $6 \times 6 \times 6$ k-point grid and an energy convergence threshold of $10^{-16}$ Ry. Finally, the electron-phonon matrix elements are interpolated onto a fine $48 \times 48 \times 48$ k-point grid. The electron-phonon calculations have been converged with respect to the broadening parameter $\sigma$ of the double-delta approximation with values of $\sigma$ ranging from 0.2~Ry to 0.5~Ry depending on the system under consideration.

\section{Results}

\begin{figure}[!tb]
\centering
\includegraphics[]
{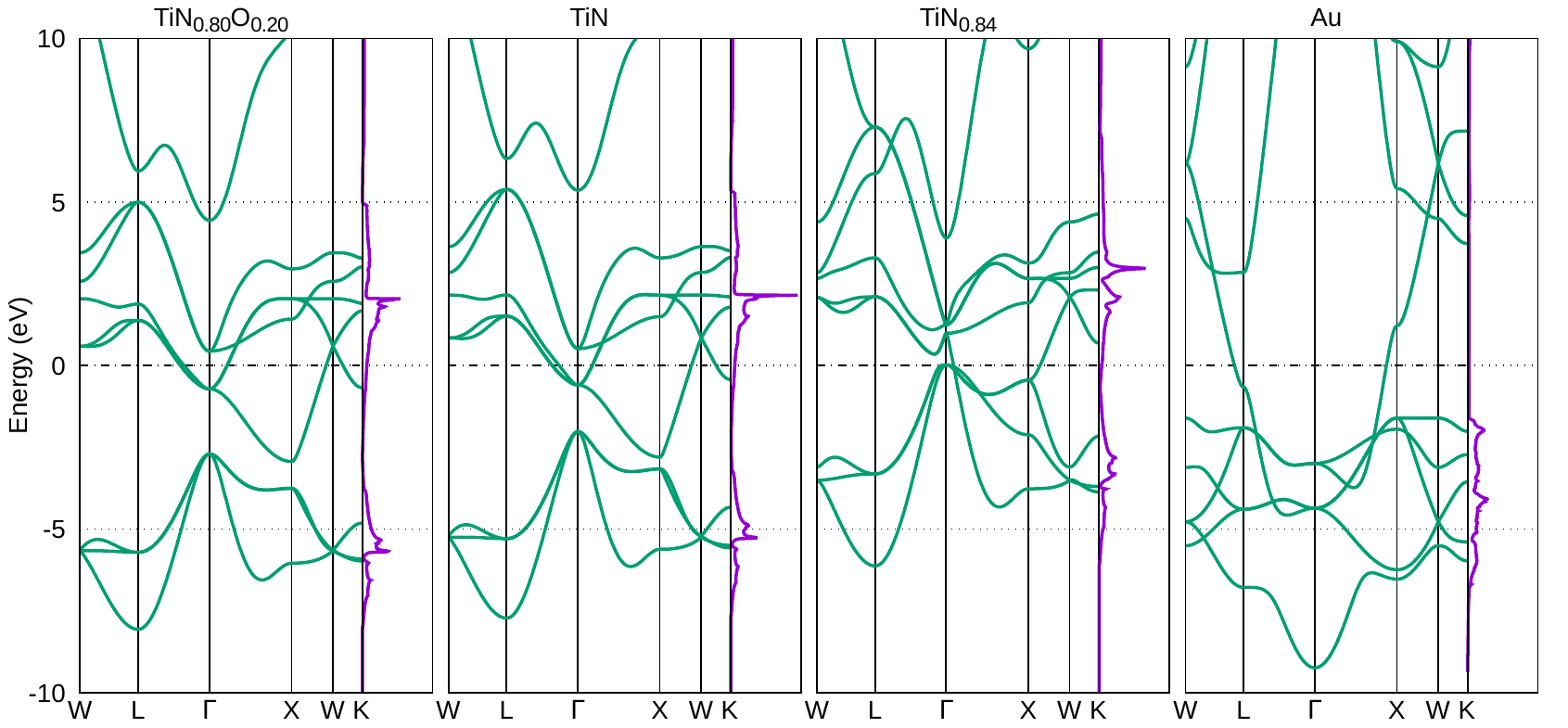}
\caption[\ch{TiN_{0.80}O_{0.20}}, \ch{TiN}, \ch{TiN_{0.84}} and \ch{Au} electron band structure]{Electron band structure and density of states of \ch{TiN_{0.80}O_{0.20}}, \ch{TiN}, \ch{TiN_{0.84}} and \ch{Au}. The zero of the energy is set to the Fermi level.}
\label{fig:elbands}
\end{figure}

Figure \ref{fig:elbands} shows the calculated electronic band structures and densities of states of \ch{TiN_{0.80}O_{0.20}}, \ch{TiN}, \ch{TiN_{0.84}} and Au. The results for the other stoichiometries are provided in the supplementary information. Pristine \ch{TiN} has a metallic character with several bands crossing the Fermi level near the $\Gamma$ and $W$ points. The density of states at the Fermi level is found to be 0.92 states/eV per unit cell. Considering \ch{TiN_{0.80}O_{0.20}}, we find that oxygen substitutionals increase the electron density and shift the Fermi level to a higher energy. The density of states at the Fermi level increases to 1.0 states/eV per unit cell. In contrast, the presence of nitrogen vacancies in \ch{TiN_{0.84}} reduces the electron density and leads to significant qualitative changes in the band structure. The electron density of states at the Fermi level is reduced to 0.57 states/eV per unit cell. For comparison, the gold band structure shows bands crossing the Fermi level near the $L$ and $X$ points and a DOS of 0.30 states/eV per unit cell at the Fermi level.

\begin{figure}[!tb]
\centering
\includegraphics[]{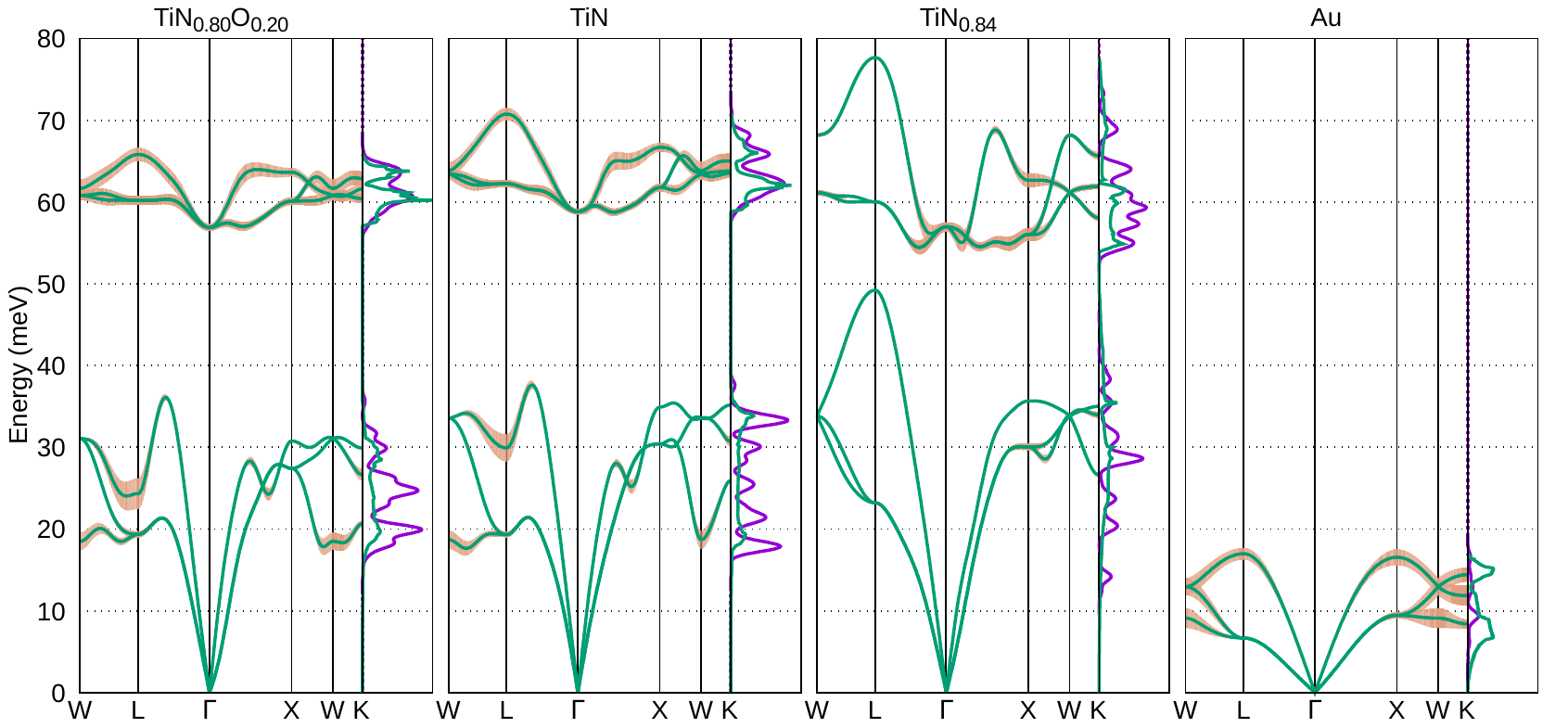}
\caption[\ch{TiN_{0.80}O_{0.20}}, \ch{TiN}, \ch{TiN_{0.84}} and \ch{Au} phonon band structure]{Phonon band structure, density of states and Eliashberg function of \ch{TiN_{0.80}O_{0.20}}, \ch{TiN}, \ch{TiN_{0.84}} and \ch{Au}. The shaded areas are proportional to the phonon linewidth $\gamma_{\mathbf{q}\nu}$ in meV: for plotting purposes, the linewidths of \ch{TiN} systems were multiplied by a factor of 2, for gold with a factor of 200.}
\label{fig:phbands}
\end{figure}

Figure~\ref{fig:phbands} shows the phonon dispersion relations and the corresponding phonon densities of states of \ch{TiN_{0.80}O_{0.20}}, \ch{TiN}, \ch{TiN_{0.84}} and Au. Again, results for the other stoichiometries are provided in the supplementary materials. For pristine \ch{TiN}, the highest phonon frequency is found to be $\sim 70$~meV corresponding to a Debye temperature of $\sim 900$~K. Comparing to the defective systems, we find that oxygen substitutionals lower the Debye temperature, while nitrogen vacancies increase it. Also shown are the phonon linewidths due to electron-phonon interactions. We find that the acoustic modes in \ch{TiN} and \ch{TiN_{0.80}O_{0.20}} exhibit large linewidths near the $W$ and $L$ points. The optical phonon bands possess a finite linewidth throughout the entire Brillouin zone indicating that these modes interact strongly with electrons. In contrast, the nitrogen-deficient system \ch{TiN_{0.84}} exhibits smaller linewidths indicating weaker electron-phonon coupling. For comparison, we obtain a Debye temperature of $\sim 165$~K for Au in excellent agreement with previous works \cite{bonn_ultrafast_2000}. Phonon linewidths in Au are two orders of magnitude smaller than in \ch{TiN} (note that the linewidths of Au in Fig.~\ref{fig:phbands} were increased by a factor of 100 compared to \ch{TiN} for plotting).

\begin{figure}[!tb]
\centering
\includegraphics[]
{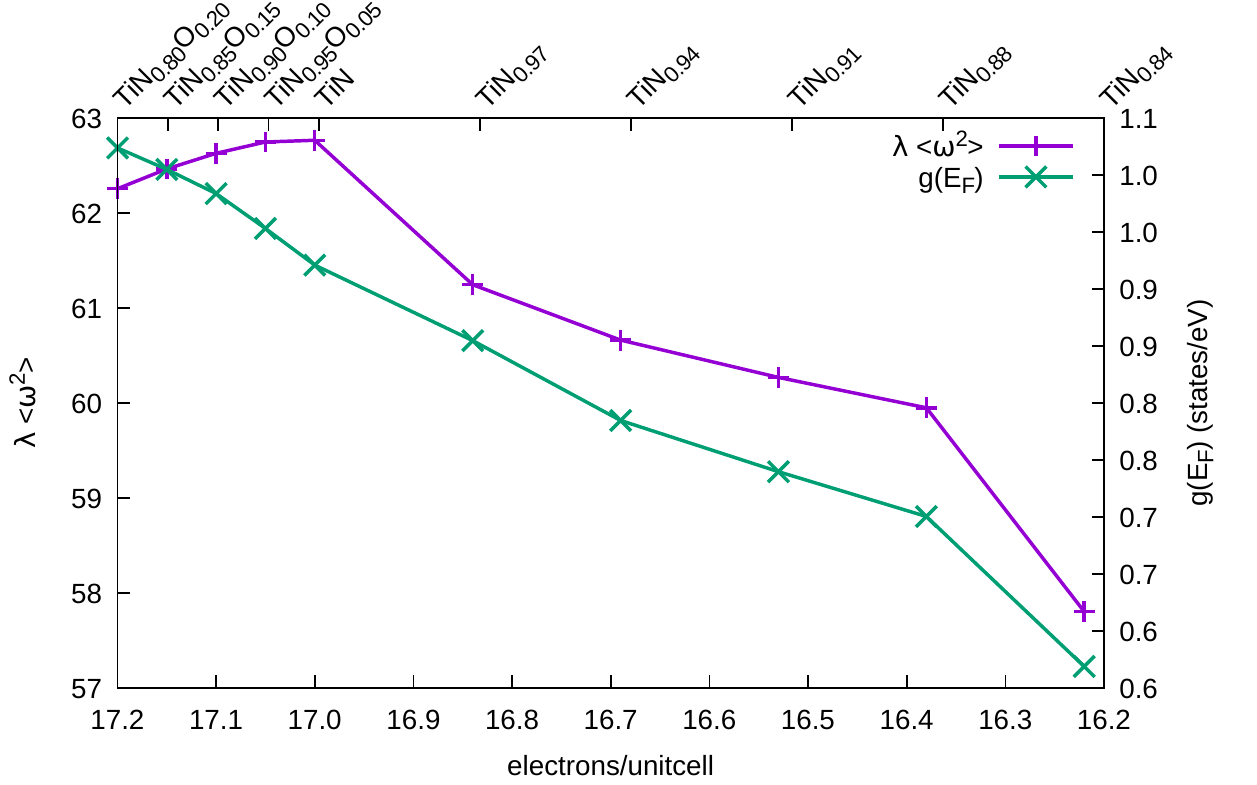}
\caption[Lambda omega square for different \ch{TiN} systems]{Second moment of the Eliashberg function and electron density of states at the Fermi level of \ch{TiN} with different defect concentrations as a function of the number of electrons in the unit cell.}
\label{fig:lambda}
\end{figure}

Figure~\ref{fig:lambda} shows the second moment of the Eliashberg function $\lambda \langle \omega^2 \rangle$ and the density of states of states at the Fermi level $g_e(E_F)$ for pristine and defective \ch{TiN}. As the concentration of nitrogen vacancies increases, both $\lambda \langle \omega^2 \rangle$ and $g(E_F)$ decrease. In contrast, as the concentration of oxygen substitutionals increases, the value of $\lambda \langle \omega^2 \rangle$ decreases while $g_e(E_F)$ is rising. Another effect that influences $\lambda \langle \omega^2 \rangle$ is the change of the average atomic mass in defective samples as the electron-phonon matrix elements are proportional to the inverse square root of the atomic mass. However, we find that the changes in $\lambda \langle \omega^2 \rangle$ are significantly larger than estimates based on the inverse mean atomic mass. This suggests that changes in $g(E_F)$ are the dominant factor that determines $\lambda \langle \omega^2 \rangle$ in defective TiN.

\begin{figure}[!tb]
\centering
\includegraphics[]{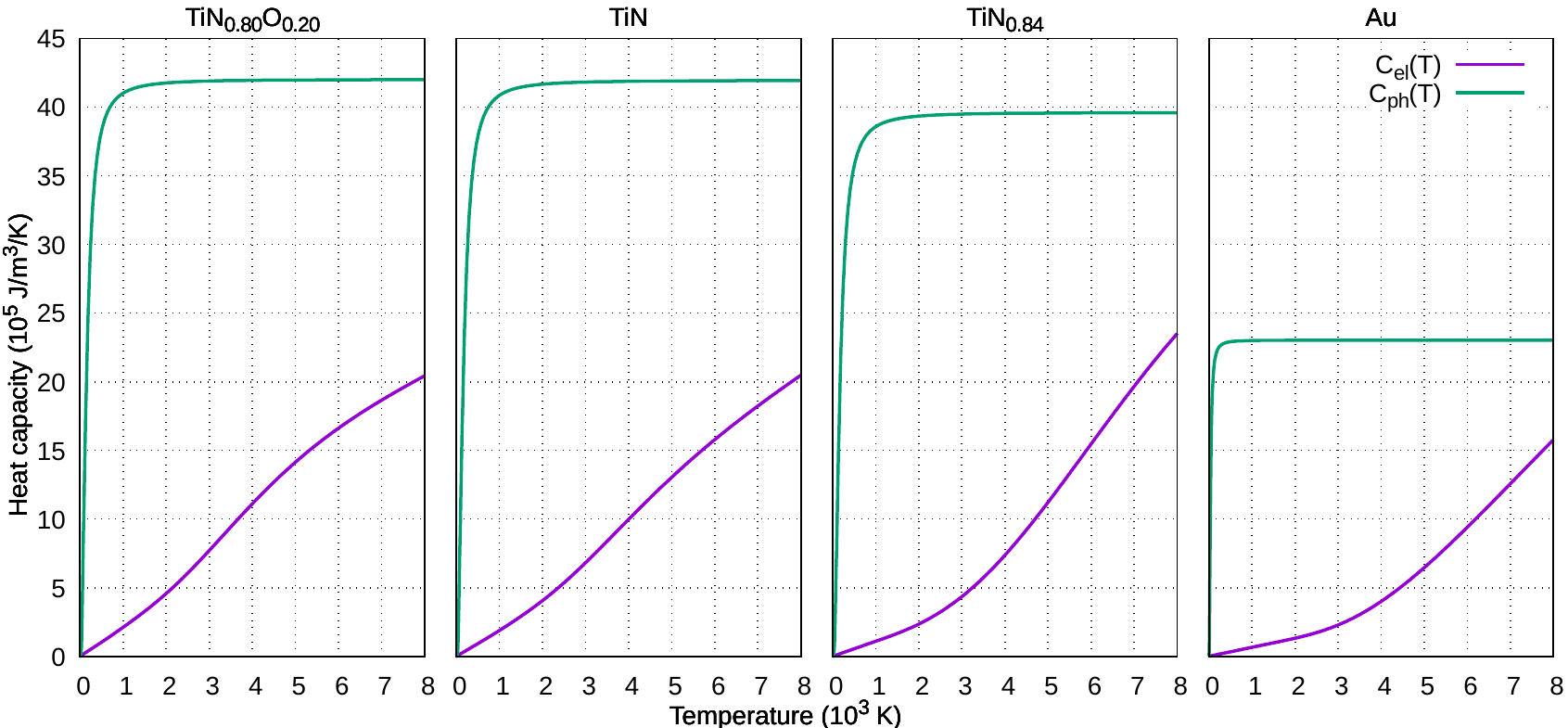}
\caption[\ch{TiN_{0.80}O_{0.20}}, \ch{TiN}, \ch{TiN_{0.84}} and \ch{Au} electron and phonon heat capacities]{Electron and phonon heat capacities of \ch{TiN_{0.80}O_{0.20}}, \ch{TiN}, \ch{TiN_{0.84}} and \ch{Au} as function of  temperature.}
\label{fig:heat}
\end{figure}

Figure~\ref{fig:heat} shows the electron and phonon heat capacities for pristine \ch{TiN}, \ch{TiN_{0.80}O_{0.20}}, \ch{TiN_{0.84}} and gold as function of temperature. As expected of metals, the electron heat capacity increases linearly at low temperatures for all systems. The phonon specific heat is proportional to $T^3$ at low temperatures and reaches a constant value when the temperature is larger than the Debye temperature. Comparing the pristine and defective \ch{TiN} systems, we observe only relatively small differences: while the electron and phonon specific heats of \ch{TiN} and \ch{TiN_{0.80}O_{0.20}} are almost identical, \ch{TiN_{0.84}} exhibits a slightly higher electron specific heat and a lower phonon specific heat. Even at high temperatures, the phonon specific heat is significantly larger than the electron specific heat for all systems. As a consequence, the phonon contribution to the electron thermalization time, see Eq.~\eqref{eq:2TM4}, is much smaller than the electron contribution and $\tau_{ep}$ becomes a function of the electron temperature only~\cite{allen_theory_1987,gadermaier_electron-phonon_2010}.

\begin{figure}[!tb]
\centering
\includegraphics[]
{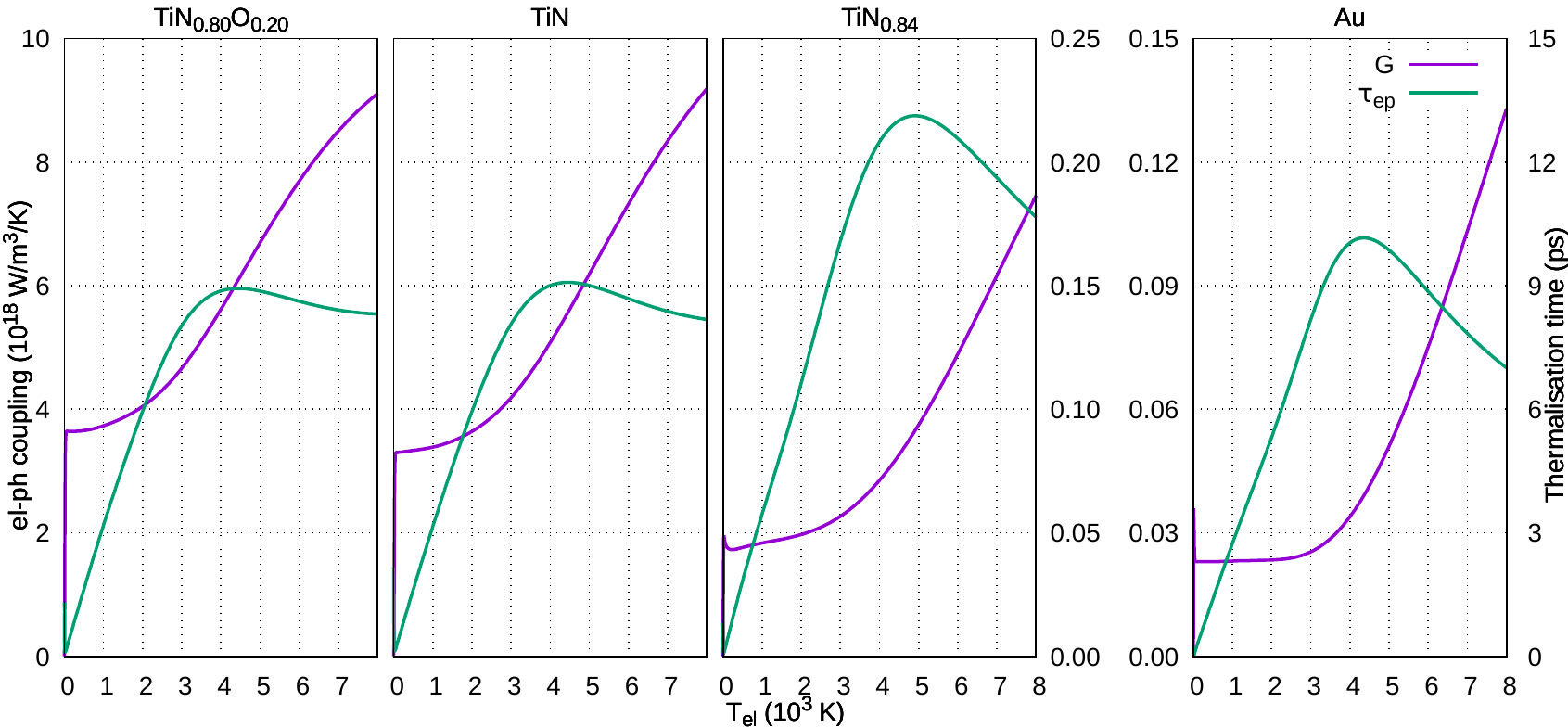}
\caption[\ch{TiN_{0.80}O_{0.20}}, \ch{TiN}, \ch{TiN_{0.84}} and \ch{Au} electron-phonon coupling parameter and thermalization time]{Electron-phonon coupling parameter $G$ and hot electron thermalization times $\tau_{ep}$ of \ch{TiN_{0.80}O_{0.20}}, \ch{TiN}, \ch{TiN_{0.84}} and \ch{Au} as a function of the electron temperature.}
\label{fig:gtau}
\end{figure}

Figure~\ref{fig:gtau} shows our results for the electron-phonon coupling parameter $G$ and the electron thermalization time $\tau_{ep}$ for pristine \ch{TiN}, \ch{TiN_{0.80}O_{0.20}}, \ch{TiN_{0.84}} and gold as a function of the electron temperature. Again, the results for the other stoichiometries are provided in the supplementary information. As expected from Eq.~\eqref{eq:gapprox}, $G$ is relatively constant at low electron temperatures. \ch{TiN_{0.80}O_{0.20}} exhibits the largest coupling parameter and \ch{TiN_{0.84}} the smallest. This is in line with the trends for the electron density of states at the Fermi level and the second moments of the Eliashberg function discussed above. The thermalization times of all systems increase linearly with electron temperature as predicted by Eq.~\eqref{eq:allen} and then reach a maximum after which $\tau_{ep}$ starts to decrease again. For pristine \ch{TiN} and \ch{TiN_{0.80}O_{0.20}}, the maximum $\tau_{ep}=150$~fs is reached at $\sim 4000$~K. For \ch{TiN_{0.84}}, the smaller electron-phonon coupling parameter leads to a larger maximum thermalization time of $215$~fs at $\sim 5000$~K. For comparison, the maximum thermalization time of Au is found to be $10.2$~ps at $T_e \sim 4000$~K. This is in good agreement with pump-probe measurements using low pump fluences for Au that have observed electron thermalization times in the range of 1-10 ps~\cite{sun_femtosecond-tunable_1994,suarez_dynamics_1995}. Moreover, our low-$T_e$ value of $G=2.2 \times 10^{16} \, W/m^3/K$ and the calculated mass enhancement parameter $\lambda=0.19$ of gold are in excellent agreement with experiments and previous DFT calculations~\cite{bonn_ultrafast_2000,mueller_nonequilibrium_2014,jain_thermal_2016}. Note that the maximum thermalization time of Au is about two orders of magnitude larger than in \ch{TiN}. This is a consequence of the significantly weaker electron-phonon coupling in Au.

\begin{figure}[!tb]
\centering
\includegraphics[]
{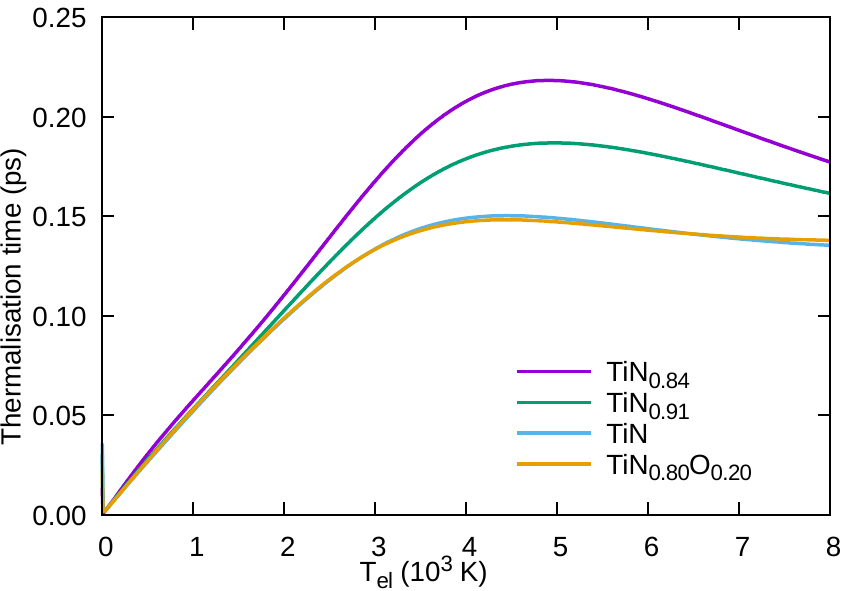}
\caption[Hot electron thermalization times of different \ch{TiN} systems]{Comparison of the hot carrier thermalization times for \ch{TiN} with different defect concentrations as a function of the electron temperature.}
\label{fig:taus}
\end{figure}

Figure~\ref{fig:taus} compares the electron thermalization time of pristine \ch{TiN} with defective samples. At low electron temperatures, the thermalization times of all systems are quite similar. At electron temperatures larger than $\sim 2000$~K, clear differences between the various systems can be observed. Again, it can be seen that the introduction of oxygen substitutionals results in almost no change compared to pristine \ch{TiN}. In contrast, nitrogen vacancies result in longer thermalization times. As discussed above, this is a consequence of the reduced electron-phonon coupling strength.

\section{Conclusions}

We have calculated hot carrier thermalization times due to electron-phonon interactions in both pristine and defective \ch{TiN} systems. To validate our approach, we also carried out calculations for Au finding excellent agreement with previous theoretical and experimental work. The thermalization times of hot carriers in \ch{TiN} are significantly shorter than in Au. Specifically, for defect-free \ch{TiN} we obtain a maximum thermalization time of 0.15~ps, at least one order of magnitude smaller than in gold. While substitution of nitrogen atoms by oxygen atoms does not have a significant influence on thermalization times, introducing nitrogen vacancies increases the maximum thermalization time to 0.215~ps. Our findings will play an important role for the design of hot carrier devices based on titanium nitride via defect engineering. 

\begin{suppinfo}
DFT results for gold and all the \ch{TiN} systems considered: \ch{Au}, \ch{TiN_{0.80}O_{0.20}}, \ch{TiN_{0.85}O_{0.15}}, \ch{TiN_{0.90}O_{0.10}}, \ch{TiN_{0.95}O_{0.05}}, \ch{TiN_{1.00}}, \ch{TiN_{0.97}}, \ch{TiN_{0.94}}, \ch{TiN_{0.91}}, \ch{TiN_{0.88}}, \ch{TiN_{0.84}}.
\ch{TiN} systems have been ordered with respect to the number of electrons per unit cell.
\end{suppinfo}

\begin{acknowledgement}
S.D.F. and J.L. acknowledge support from EPRSC under Grant No. EP/N005244/1 and also from the Thomas Young Centre under Grant No. TYC-101. Via J.L.'s membership of the UK's HEC Materials Chemistry Consortium, which is funded by EPSRC (EP/L000202), this work used the ARCHER UK National Supercomputing Service.
\end{acknowledgement}

\bibliographystyle{achemso}
\bibliography{TiN}

\end{document}